\documentclass[final,authoryear,5p,times,twocolumn]{elsarticle}
\biboptions{square,comma,numbers,sort&compress}
\journal{Physica Scripta}

\usepackage[a4paper,ps2pdf,colorlinks=true,breaklinks=true]{hyperref}
\usepackage{bm}

\graphicspath{{fig/}}

\newcommand{\EQ}{\begin{equation}}
\newcommand{\EN}{\end{equation}}
\newcommand{\EQA}{\begin{eqnarray}}
\newcommand{\ENA}{\end{eqnarray}}

\newcommand{\Eq}[1]{equation~(\ref{#1})}

\newcommand{\Sec}[1]{\S\ref{#1}}

\newcommand{\Fig}[1]{Fig.~\ref{#1}}

\newcommand{\bra}[1]{\langle #1\rangle}

\newcommand{\meanrho}{\overline{\rho}}

{}
{}

{}

{}
{}
{}
{}
{}
{}
{}
{}
\newcommand{\meanBB}{\overline{\mbox{\boldmath $B$}}{}}{}
\newcommand{\meanJJ}{\overline{\mbox{\boldmath $J$}}{}}{}
\newcommand{\meanUU}{\overline{\mbox{\boldmath $U$}}{}}{}
{}
{}

\newcommand{\meanB}{\overline{B}}

\newcommand{\meanU}{\overline{U}}

\newcommand{\meanJ}{\overline{J}}

{}
{}
{}
{}

\def\qp{q_{\rm p}}
\def\qs{q_{\rm s}}
\def\qg{q_{\rm g}}
\def\qJ{q_{\rm J}}
\def\qI{q_{\rm I}}
\def\qF{q_{\rm F}}
\def\peff{p_{\rm eff}}
\newcommand{\uu}{\mbox{\boldmath $u$} {}}
\newcommand{\UU}{\mbox{\boldmath $U$} {}}

\newcommand{\bb}{\mbox{\boldmath $b$} {}}

\newcommand{\BB}{\mbox{\boldmath $B$} {}}

\newcommand{\JJ}{\mbox{\boldmath $J$} {}}

\newcommand{\AAA}{\mbox{\boldmath $A$} {}}

\newcommand{\ff}{\mbox{\boldmath $f$} {}}

\newcommand{\grav}{\mbox{\boldmath $g$} {}}
\newcommand{\nab}{\mbox{\boldmath $\nabla$} {}}

\newcommand{\SSSS}{\mbox{\boldmath ${\sf S}$} {}}

\newcommand{\DD}{{\rm D} {}}

\newcommand{\const}{{\rm const}  {}}

\def\Pm{\mbox{\rm Pr}_{\rm M}}
\def\Rm{\mbox{\rm Re}_{\rm M}}
\def\Rey{\mbox{\rm Re}}

\def\cs{c_{\rm s}}
\def\kf{k_{\rm f}}
\def\urms{u_{\rm rms}}

\def\etatz{\eta_{\rm t0}}

\def\Beq{B_{\rm eq}}
\def\Beqz{B_{\rm eq0}}

\def\half{{\textstyle{1\over2}}}

\def\onethird{{\textstyle{1\over3}}}

\def\betap{\beta_{p}}
\def\betacrit{\beta_{\rm crit}}
\def\betastar{\beta_{\star}}
\def\Peff{{\cal P}_{\rm eff}}

\newcommand{\etal}{{\em et al.}}
\newcommand{\yapj}[3]{ #1, {\it Astrophys.\ J.} {\bf #2} #3}

\newcommand{\papjl}[2]{ #1, {\it Astrophys.\ J.\ Lett.}, in press, arXiv:#2}

\newcommand{\yan}[3]{ #1, {\it Astron.\ Nachr.} {\bf #2} #3}

\newcommand{\yana}[3]{ #1, {\it Astron. Astrophys.} {\bf #2} #3}

\newcommand{\ypfb}[3]{ #1, {Phys.\ Fluids B,} {#2}, #3}

\newcommand{\ysovl}[3]{ #1, {\it Sov.\ Astron.\ Lett.} {\bf #2} #3}
\newcommand{\yjetp}[3]{ #1, {\it Sov.\ Phys.\ JETP} {\bf #2} #3}

\newcommand{\ymn}[3]{ #1, {\it Mon.\ Not.\ R.\ Astron.\ Soc.}{\bf #2} #3}

\newcommand{\ysph}[3]{ #1, {\it Solar Phys.} {\bf #2} #3}
\newcommand{\psph}[3]{ #1, {\it Solar Phys.} {#2}, #3}

\newcommand{\ypre}[3]{ #1, {\it Phys.\ Rev.\ E }{\bf #2} #3}

\newcommand{\yjour}[4]{ #1, {\it #2}, {#3} #4}

\newcommand{\ybook}[3]{ #1, {#2} (#3)}

\newcommand{\sana}[1]{ #1, {\it Astron. Astrophys.} (submitted)}

\newlength{\shortwidth}
\newlength{\textwidtha}
\begin{document}

\begin{frontmatter}

\title{Nonuniformity effects in the negative effective magnetic pressure instability}

\author{
\textbf{K Kemel}$^{1,2}$,
\textbf{A Brandenburg}$^{1,2}$,
\textbf{N Kleeorin}$^{3,1}$, and
\textbf{I Rogachevskii}$^{3,1}$
}

\address{
$^1$NORDITA, Roslagstullsbacken 23, SE-10691 Stockholm, Sweden\\
$^2$Department of Astronomy, Stockholm University, SE-10691 Stockholm, Sweden\\
$^3$Department of Mechanical Engineering, Ben-Gurion University of the Negev,\\
POB 653, Beer-Sheva 84105, Israel\\[2mm]
\href{mailto:kemel@nordita.org}{kemel@nordita.org, $ $Revision: 1.38 $ $}
}

\begin{abstract}
Using direct numerical simulations (DNS) and mean-field simulations (MFS),
the effects of non-uniformity of the magnetic field on the suppression of the
turbulent pressure is investigated.
This suppression of turbulent pressure can lead to an instability which, in turn,
makes the mean magnetic field even more non-uniform.
This large-scale instability is caused by a resulting negative
contribution of small-scale turbulence
to the effective (mean-field) magnetic pressure.
We show that enhanced mean current density increases the suppression of the
turbulent pressure.
The instability leads to magnetic flux concentrations in which the normalized
effective mean-field pressure is reduced to a certain value at all depths
within a structure.
\end{abstract}

\end{frontmatter}

\section{Introduction}

The Sun's magnetic field is generally believed to be
due to a turbulent dynamo operating in the convection zone
in its outer 30\% by radius \citep{KR80,Mof78,Par79,ZRS83,BS05}.
Recent simulations performed by a number of groups
have shown that the magnetic field is
produced in the bulk of the convection zone.
According to the flux tube scenario, most of the toroidal
magnetic field resides at the bottom of the convection zone,
or possibly just beneath it \citep{GD00,PM07}.
Another possibility is that most of the toroidal field resides in
the bulk of the convection zone, but that its spatio-temporal
properties are strongly affected by the near-surface dynamics
\cite{KMB12}, or the near-surface shear layer \citep{B05}.
In any case, the question then emerges how one can explain
the formation of active regions out of which sunspots develop
during the lifetime of an active region.

In the past, this question was conveniently bypassed by referring
to the possible presence of a strong toroidal flux belt at the
bottom of the convection zone, where they would be in a stable state,
except that every now and then they would become unstable, for
example to the clamshell or tipping instabilities \citep{Cally}.
However, if the magnetic field is continuously being destroyed
and regenerated by the turbulence in the convection zone proper,
the mechanism for producing active regions and eventually sunspots
must be one that is able to operate within a turbulent environment.
One such mechanism may be the negative effective magnetic pressure
instability (NEMPI) which is based on the suppression of turbulent
pressure by a weak mean magnetic field, leading therefore to a negative
{\it effective} (or mean-field) magnetic pressure and, under suitable
conditions, to an instability \citep{KRR89,KRR90,KR94,KMR96,RK07}.
This has been the subject of intensive research in recent years
\citep{BKR10,BKKR12,KBKR12,KBKMR12a,KBKMR12b,KBKMR12,Los12},
following the first detection of such
an instability in direct numerical simulations \citep[DNS; see][]{BKKMR}.
Another mechanism that has been discussed in connection with the
production of magnetic flux concentrations is related to the suppression
of the turbulent convective heat flux \citep{KM00}.
Meanwhile, simulations of realistic solar convection have demonstrated
that large-scale magnetic flux inhomogeneities can develop when horizontal
magnetic flux is injected at the bottom of the simulation domain \citep{SN12},
but it remains to be seen whether this is connected with any of the two
aforementioned mechanisms.

The purpose of the present paper is to investigate the possibility that
higher-order contributions (involving higher spatial derivatives of the
mean magnetic field) might play a role in NEMPI.
We do this by using DNS to measure the resulting turbulent transport
coefficients in cases where a measurable mean current density
develops in the DNS.
Note that even for an initially uniform mean magnetic field,
a mean current density develops as a consequence of NEMPI
itself, which redistributes an initially uniform magnetic field into a
structured one.
To investigate this process further, we use appropriately tailored
mean-field simulations (MFS) that show how the spatial variations of the
negative effective magnetic pressure vary in space as the instability
runs further into saturation.
We begin by discussing first the basic equations and turn then to the results.
Throughout this work we use an isothermal equation of state which yields the
simplest possible system to investigate this process.

\section{Basic equations}

In this paper we use both DNS and MFS to study
the effects of nonuniformity of the magnetic
field of the development of NEMPI. In the DNS,
the solutions turn out to have a large-scale
two-dimensional pattern which can best be
isolated using averaging over the $y$ direction.
Furthermore, by imposing a uniform magnetic
field, we can determine some of the turbulent
transport coefficients that characterize the
dependence of the Reynolds and Maxwell stress on
the mean field. This is generally done by
determining the total mean stress
\EQ
\overline\Pi_{ij}=\meanrho\,\overline{U_iU_j}
+\half\delta_{ij}\overline{\BB^2}-\overline{B_iB_j}.
\EN
Here, the vacuum permeability is set to unity
and overbars indicate $y$ averages. This total
mean stress has contributions from the
fluctuations,
\EQ \overline\Pi_{ij}^{\rm
f}=\meanrho\,\overline{u_iu_j}
+\half\delta_{ij}\overline{\bb^2}-\overline{b_ib_j},
\label{stress0}
\EN
where $\uu=\UU-\meanUU$ and
$\bb=\BB-\meanBB$ are the departures from the
averaged fields. Here $\meanUU$ and $\meanBB$ are
the mean velocity and magnetic fields, and
$\overline{p}$ is the mean fluid pressure. This,
together with the contribution from the mean
field, namely
\EQ
\overline\Pi_{ij}^{\rm
m}=\meanrho \, \meanU_i\meanU_j
+\delta_{ij}\left(\overline{p}+\half\meanBB^2\right)
-\meanB_i\meanB_j-2\nu \meanrho \, \overline{\sf
S}_{ij},
\EN
yields the total mean stress tensor,
i.e., $\overline\Pi_{ij}=\overline\Pi_{ij}^{\rm
m}+\overline\Pi_{ij}^{\rm f}$.
The term $\overline\Pi_{ij}^{\rm m}$ depends only on the
mean field and is therefore directly obtained in
MFS, while $\overline\Pi_{ij}^{\rm f}$ is caused by the fluctuating
velocity and magnetic fields and requires a parameterization.
It has a contribution independent of the mean field,
$\overline{\Pi}^{\rm f,0}_{ij}$, and one that depends on it,
$\Delta\overline\Pi_{ij}^{\rm f}(\meanB)$.
Much of the recent work in this field focussed on the
parameterization
\EQ
\Delta\overline\Pi_{ij}^{\rm f}=
-\half\qp(\beta)\, \delta_{ij}\, \meanBB^2 + \qs(\beta) \, \meanB_i \meanB_j
+ \qg(\beta) \, \meanBB^2 \hat g_i \hat g_j,
\label{qg}
\EN
where $\hat g_i=g_i/g$ is the unit vector in the direction of gravity.
This difference in the mean stress, $\Delta\overline\Pi_{ij}^{\rm f}(\meanB)$,
is caused solely by the presence of the mean magnetic field $\meanBB$,
where $\qp(\beta)$ is found to be a positive
function of $\beta=|\meanBB|/\Beq$ only, and, for weak magnetic fields,
$\qp(\beta)$ is well in excess of unity for $\Rm\gg1$, thus overcoming
the magnetic pressure from the mean field itself.
However the functions $\qs(\beta)$ and $\qg(\beta)$ were found to be small for
isothermal turbulence.
The net result
for the sum $\overline\Pi_{ij}^{\rm m}+\Delta\overline\Pi_{ij}^{\rm f}$ is
\EQ
\overline\Pi_{ij} \approx \overline{\Pi}^{\rm f,0}_{ij} +
\delta_{ij} \, \peff(\meanB/\Beq) - \meanB_i \meanB_j,
\EN
where $\peff=\half\left[1-\qp(\beta)\right] \, \meanBB^2$
is the mean effective magnetic pressure
that is negative for $\beta<\betacrit$, where $\betacrit\approx0.5$
(depending on other details of the system).
This results in a large-scale instability (NEMPI)
and the formation of large-scale inhomogeneous magnetic structures.

In the nonlinear stage of NEMPI, the mean magnetic field becomes strongly
nonuniform.
This implies that the Maxwell--Reynolds stress tensor
$\Delta\overline\Pi_{ij}^{\rm f}$ may depend also on spatial derivatives of
the mean magnetic field, i.e.,
\begin{eqnarray}
\Delta\overline\Pi_{ij}^{\rm f}&=&-\half\delta_{ij}\, \qp\, \meanBB^2
+ \qs \,  \meanB_i \meanB_j + \qg \, \meanBB^2 \hat g_i \hat g_j
+ C_1 \meanB_{i,m}\meanB_{j,m}
\nonumber\\
&&+ C_2 \meanB_{m,i}\meanB_{m,j} + C_3\, (\meanB_{i,m}\meanB_{m,j}+
\meanB_{j,m}\meanB_{m,i}),
\label{B1}
\end{eqnarray}
where $\meanB_{i,j}=\nabla_j \meanB_{i}$. We decompose $\meanB_{i,j}$
into symmetric and antisymmetric parts:
\begin{eqnarray}
\meanB_{i,j}=(\partial \meanB)_{ij}-\half \varepsilon_{ijm} \meanJ_{m},
\label{B2}
\end{eqnarray}
where $(\partial \meanB)_{ij}=\half (\meanB_{i,j}+\meanB_{j,i})$.
Substituting \Eq{B2} into \Eq{B1} we obtain:
\begin{eqnarray}
\Delta\overline\Pi_{ij}^{\rm f}&=&-\half\delta_{ij}\, \qp\, \meanBB^2
+ \qs \,  \meanB_i \meanB_j + \qg \, \meanBB^2 \hat g_i \hat g_j
\nonumber\\
&& - \qJ \, \left(\meanJ^2 \delta_{ij} - \meanJ_{i}\meanJ_{j}\right)
- \qF \, (\partial \meanB)_{im} (\partial \meanB)_{mj}
\nonumber\\
&&- \qI \, \left(\varepsilon_{iml} (\partial \meanB)_{mj} +
\varepsilon_{jml} (\partial \meanB)_{mi}\right) \meanJ_{l} .
\label{B3}
\end{eqnarray}
Let us consider a mean magnetic field of the form
$\meanBB=(0, \meanB_y(x,z), 0)$, so
$(\partial \meanB)_{xy} =\meanJ_z/2$, $(\partial \meanB)_{yz} =-\meanJ_x/2$ and
\begin{eqnarray}
(\partial \meanB)_{im} (\partial \meanB)_{mj} =
{1\over4}\left( \matrix { \meanJ_z^2 & 0 & - \meanJ_x \meanJ_z
\cr
0 & \meanJ^2 & 0
\cr
- \meanJ_x \meanJ_z & 0 & \meanJ_x^2
\cr } \right) ,
\label{B4}
\end{eqnarray}
\begin{eqnarray}
&&\left(\varepsilon_{iml} (\partial \meanB)_{mj} +
\varepsilon_{jml} (\partial \meanB)_{mi}\right) \meanJ_{l}
\nonumber\\
&& \quad \quad \quad \quad =
{1\over2}\left( \matrix { \meanJ_z^2 & 0 & - \meanJ_x \meanJ_z
\cr
0 & -\meanJ^2 & 0
\cr
- \meanJ_x \meanJ_z & 0 & \meanJ_x^2
\cr } \right) .
\label{BB4}
\end{eqnarray}

Equations~(\ref{B3}) and~(\ref{B4}) yield:
\begin{eqnarray}
\Delta\overline\Pi_{xx}^{\rm f} &=& - \tilde \qJ \, \meanJ_z^2
- \half \qp \, \meanB^2,
\label{B5}\\
\Delta\overline\Pi_{yy}^{\rm f} &=& \tilde \qI \, \meanJ^2
+ \left(\qs- \half\qp\right) \, \meanB^2 ,
\label{B6}\\
\Delta\overline\Pi_{zz}^{\rm f} &=& - \tilde \qJ \, \meanJ_x^2
+ \left(\qg- \half\qp\right) \, \meanB^2,
\label{B7}\\
\Delta\overline\Pi_{xz}^{\rm f} &=& \tilde \qJ \, \meanJ_x
\, \meanJ_z,
\label{B8}
\end{eqnarray}
where $\tilde \qJ(\meanB^2, \meanJ^2)=\qJ+\qI/2+\qF/4$,~
$\tilde \qI(\meanB^2, \meanJ^2)=\qI/2-\qJ-\qF/4$,
$\qp=\qp(\meanB^2, \meanJ^2)$,
$\qs=\qs(\meanB^2, \meanJ^2)$ and $\qg=\qg(\meanB^2, \meanJ^2)$.
Unfortunately, we have only 4 equations, but 5 unknowns, so we cannot
obtain all the required transport coefficients independently.
In the following, we can only draw some limited conclusions that will
allow us to motivate a numerical assessment of the nonlinear
$(\meanB^2, \meanJ^2)$ dependence of $\qp$.

\section{Results}

\subsection{DNS model}
\label{model}

Following the earlier work of \cite{BKKMR} and \cite{KBKMR12a,KBKMR12b},
we solve the isothermal equations for the velocity, $\UU$,
the magnetic vector potential, $\AAA$, and the density, $\rho$
\begin{eqnarray}
\rho{\DD\UU\over\DD t}&=&-\cs^2\nab\rho+\JJ\times\BB+\rho(\ff+\grav)
+\nab\cdot(2\nu\rho\SSSS),\\
{\partial\AAA\over\partial t}&=&\UU\times\BB+\eta\nabla^2\AAA,\\
{\partial\rho\over\partial t}&=&-\nab\cdot\rho\UU,
\end{eqnarray}
where $\nu$ is the kinematic viscosity, $\eta$ is the magnetic diffusivity
due to Spitzer conductivity of the plasma,
$\BB=\BB_0+\nab\times\AAA$ is the magnetic field,
$\BB_0=(0,B_0,0)$ is the imposed uniform field,
$\JJ=\nab\times\BB/\mu_0$ is the current density,
$\mu_0$ is the vacuum permeability, ${\sf S}_{ij}
=\half(U_{i,j}+U_{j,i})-\onethird\delta_{ij}\nab\cdot\UU$
is the traceless rate-of-strain tensor.
The forcing function, $\ff$, consists of random, white-in-time,
plane, non-polarized waves with a certain average wavenumber, $\kf$.
The turbulent rms velocity is approximately
independent of $z$ with $\urms=\bra{\uu^2}^{1/2}\approx0.1\,\cs$, where
$\cs=\const$ is the isothermal sound speed.
The gravitational acceleration, $\grav=(0,0,-g)$ is chosen such that
$k_1 H_\rho=1$, so the density contrast between
bottom and top is $\exp(2\pi)\approx535$.
Here, $H_\rho=\cs^2/g$ is the density scale height
and $k_1=2\pi/L$ is the smallest wavenumber that fits into
the cubic domain of size $L^3$.
We consider a domain of size $L_x\times L_y\times L_z$ in
Cartesian coordinates $(x,y,z)$, with periodic boundary conditions in
the $x$- and $y$-directions and stress-free, perfectly conducting
boundaries at the top and bottom ($z=\pm L_z/2$).
In the following we refer to $\kf/k_1$ as the scale separation ratio,
for which we choose the value $30$ in all cases.
For the fluid Reynolds number we take $\Rey\equiv\urms/\nu\kf=18$,
and for the magnetic Prandtl number $\Pm=\nu/\eta=0.5$.
The magnetic Reynolds number is $\Rm=\Pm\Rey$.
In our units, $\mu_0=1$ and $\cs=1$.
The simulations are performed with the {\sc Pencil Code}\footnote{
\href{http://pencil-code.googlecode.com}{http://pencil-code.googlecode.com}}
which uses sixth-order explicit finite differences in space and a
third-order accurate time stepping method.
We use a numerical resolution of $256^3$ mesh points.
In the MFS we also use $128$ meshpoints, but because the MFS are
two-dimensional, our resolution is $128^2$ mesh points.
In the model presented below, the $z$ extent is however slightly bigger:
$z/H_\rho$ is 8 instead of $2\pi$.

\subsection{DNS results}
\label{DNS}

\begin{figure}[t!]\begin{center}
\includegraphics[width=\columnwidth]{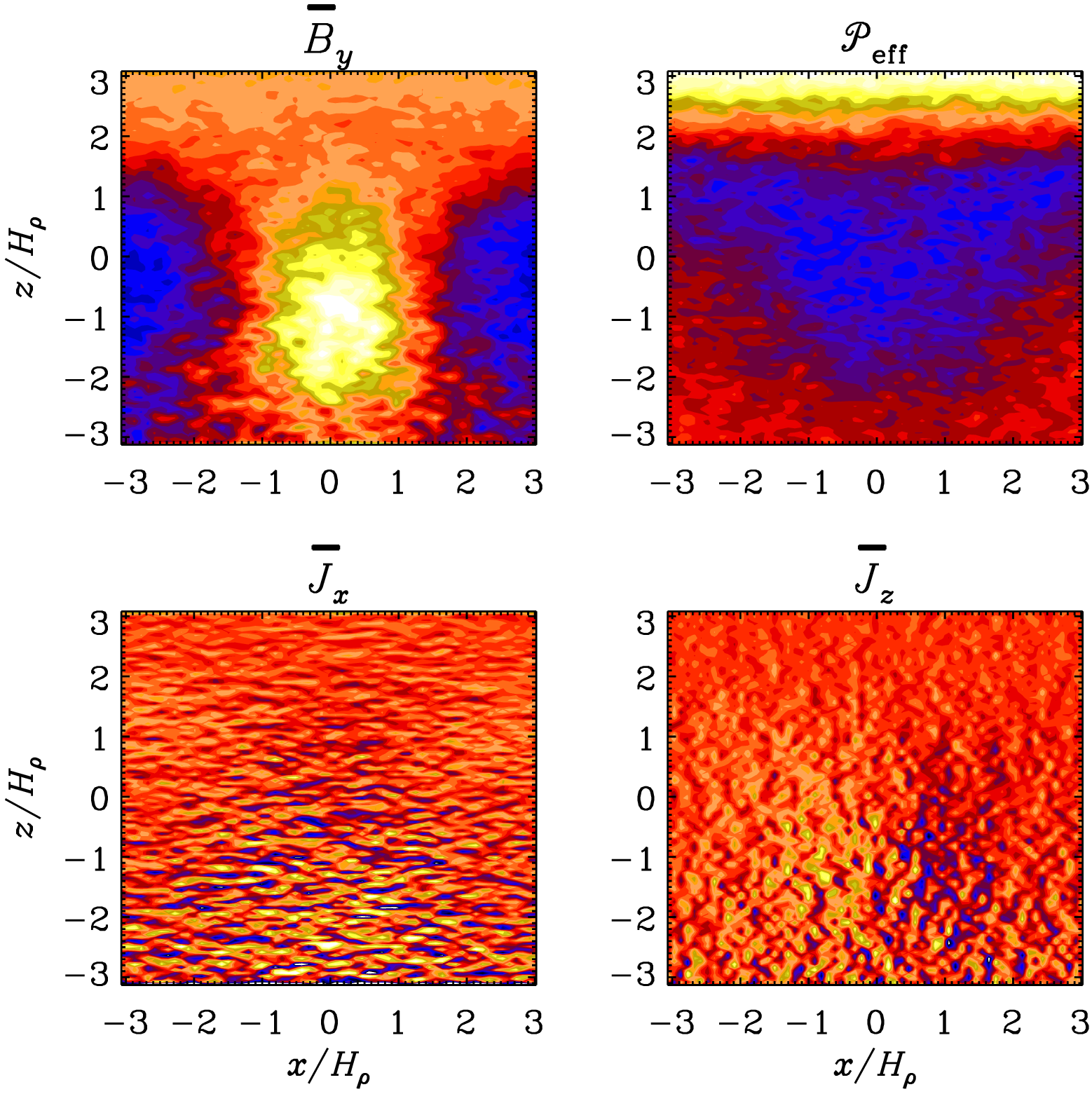}
\end{center}\caption[]{
Representations of $\meanB_y$, $\Peff$, and the two components of $\meanJJ$
in the $xz$ plane from a DNS with $\Rm=18$.
}
\label{DNS_J}
\end{figure}

In all cases, we consider a weak imposed magnetic field in the $y$ direction.
In \Fig{DNS_J} we show $y$-averaged visualizations of the normal component
of the magnetic field, $\meanB_y$, together with the two components of the
mean current density, $\meanJ_x=-\partial\meanB_y/\partial z$, and
$\meanJ_z=\partial\meanB_y/\partial x$, and the normalized effective
magnetic pressure, $\Peff=\Delta \overline\Pi_{xx}^{\rm f}/\Beq^2$.

As a result of NEMPI, the field in the $xz$ plane gets concentrated in
one position and diluted in another; see the first panel of \Fig{DNS_J}.
This leads to an equilibrium in which the resulting reduction of the
turbulent pressure is offset by a corresponding increase in the gas pressure
and therefore a corresponding increase in the density.
The nonuniformity of the magnetic field implies a non-vanishing current
density which is best seen in $\meanJ_z$ (lower right panel of \Fig{DNS_J}),
but this is mainly because of enhanced fluctuations resulting from
variations in the $z$ direction.

\begin{figure}[t!]\begin{center}
\includegraphics[width=\columnwidth]{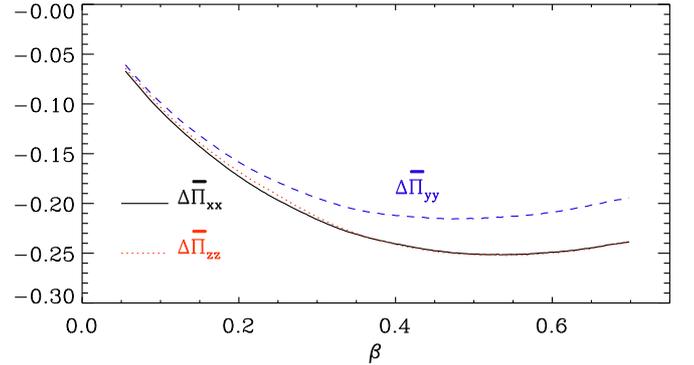}
\end{center}\caption[]{
The 3 diagonal components of the tensor
$\Delta\overline\Pi_{xx}^{\rm f}$ (solid, black),
$\Delta\overline\Pi_{yy}^{\rm f}$ (dashed, blue),
and $\Delta\overline\Pi_{zz}^{\rm f}$ (dotted, red),
normalized by $\Beq^2$ with a zero current density.
}\label{DPii}\end{figure}

\begin{figure}[t!]\begin{center}
\includegraphics[width=\columnwidth]{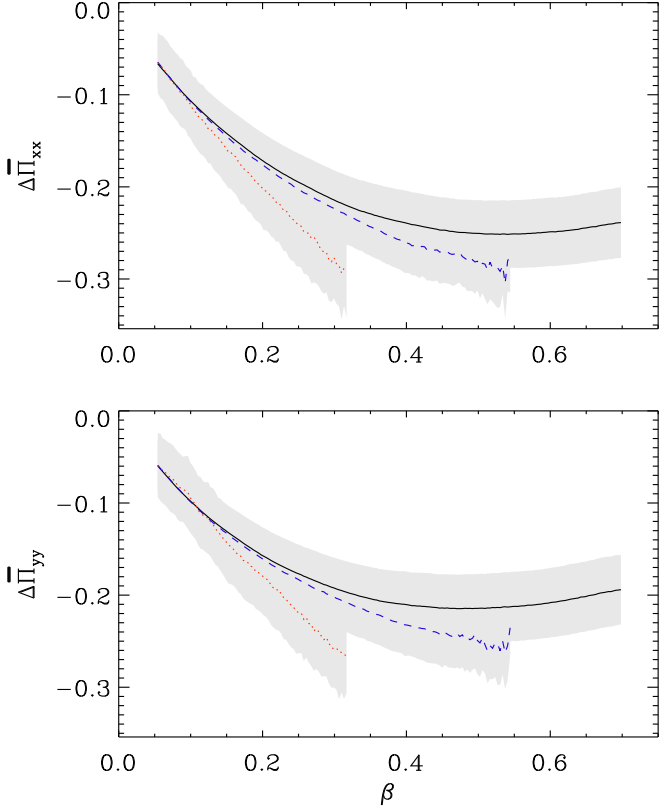}
\end{center}\caption[]{
Normalized diagonal components $\Delta\overline\Pi_{xx}^{\rm f}$ and
$\Delta\overline\Pi_{yy}^{\rm f}$ as a function of $\beta$, for
vanishing mean current density (black, solid line),
for points with low (dashed, blue, $0.25 > \meanJ^2H_{\rho}^2/B_0^2 > 0.1$)
and higher current densities (dotted, red, 0.25 < $J^2/H_{\rho}^2B_0^2$).
}\label{qp_J}\end{figure}

In \Fig{DPii} we show three diagonal components of the tensor
$\Delta\overline\Pi_{xx}^{\rm f}$, $\Delta\overline\Pi_{yy}^{\rm f}$
and $\Delta\overline\Pi_{zz}^{\rm f}$, normalized by $\Beq^2$
in turbulence with a zero mean
current density. \Fig{DPii} demonstrates that the tensor
$\Delta\overline\Pi_{yy}^{\rm f}$ in the direction of the mean magnetic field
is different from the tensors $\Delta\overline\Pi_{xx}^{\rm f}$
and $\Delta\overline\Pi_{zz}^{\rm f}$ in the directions perpendicular to
the mean magnetic field.

\begin{figure}[t!]\begin{center}
\includegraphics[width=\columnwidth]{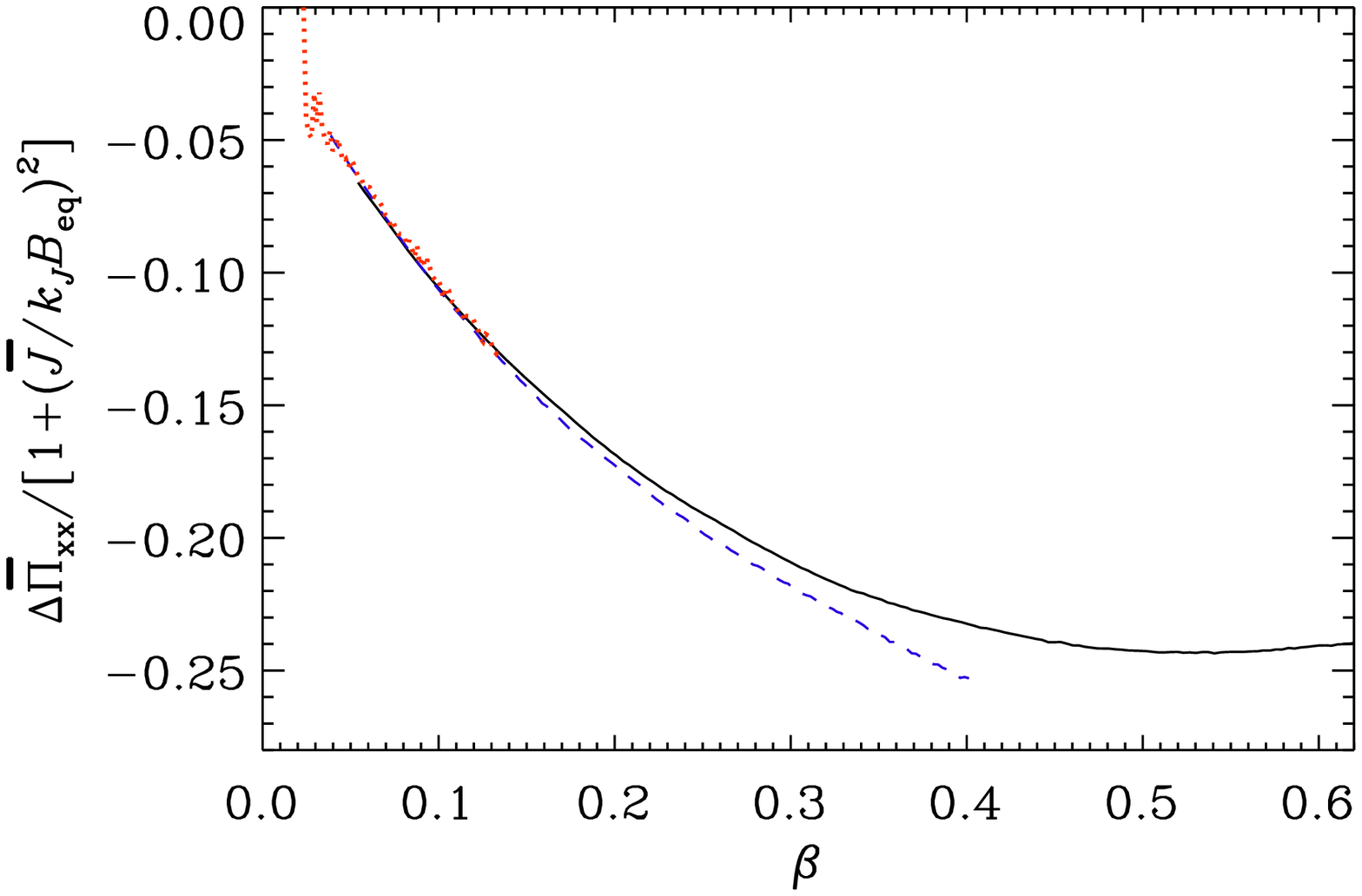}
\end{center}\caption[]{
Like \Fig{qp_J}, but compensated by $1/(1+\meanJ^2/k_J^2\Beq^2)$
for $k_J H_\rho=4$.
}\label{qp_dpii_c_fit}\end{figure}

In the following we want to study the possible effects of current density
on the resulting mean-field (or effective) magnetic pressure.
We find that $\Delta\overline\Pi_{xz}^{\rm f}$
vanishes, which implies that $\tilde \qJ = 0$; see Eq.~(\ref{B8}).
On the other hand, since $\Delta\overline\Pi_{xx}^{\rm f}
\approx \Delta\overline\Pi_{zz}^{\rm f}$, the coefficient $\qg$ vanishes
or is small.
In \Fig{qp_J} we show two diagonal components
$\Delta\overline\Pi_{xx}^{\rm f}$ and $\Delta\overline\Pi_{yy}^{\rm f}$,
normalized by $\Beq^2$ for different mean current densities.
Inspection of \Fig{qp_J} shows that
the mean current density increases the negative minimum
of the effective magnetic pressure
characterized by $\Delta\overline\Pi_{xx}^{\rm f}$.
This implies that an enhanced current density increases
the effect of negative effective magnetic pressure, i.e., they intensify
the formation of magnetic structures.
Therefore, Eq.~(\ref{B5}) allows us to determine $\qp =
-2\Delta\overline\Pi_{xx}^{\rm f} /\meanB^2$ and
the mean effective magnetic pressure
$\peff=\half\left[1-\qp(\beta)\right] \, \meanB^2
= \Delta\overline\Pi_{xx}^{\rm f}+ \half\meanB^2$.
In agreement with earlier studies \citep{BKKR12}, we find a
clear negative minimum in $\Peff(\beta)$ at $\beta\approx0.25$.
However, as the current density increases, the minimum of $\Peff(\beta)$
deepens, suggesting that
NEMPI might turn out to be stronger than originally anticipated based
on the dependence $\Peff(\beta)$ that does not distinguish between strong
and weak current densities.
A reasonable fit to such a behavior would be of the form
\EQ
\overline\Pi_{xx}=-\half(1+\meanJ^2/k_J^2\Beq^2)\qp(\beta)\,\meanB^2,
\EN
where $k_J$ is a free parameter.
In \Fig{qp_dpii_c_fit} we show that we can get a good fit to the data
for $k_J H_\rho=4$.
Note further that \Eq{B6} has two unknowns $\tilde \qI$ and
$\qs$ which cannot be determined for such a simple configuration
of the mean magnetic field, $\meanBB=(0, \meanB_y(x,z), 0)$.

\subsection{Gravity effects in DNS}
\label{Gravity}

\begin{figure}[t!]\begin{center}
\includegraphics[width=\columnwidth]{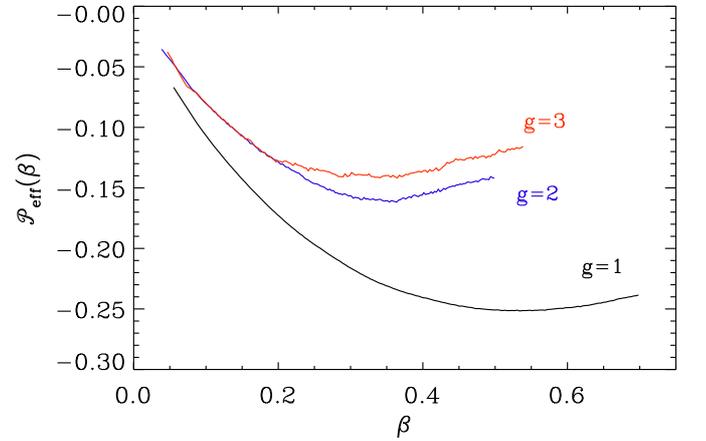}
\end{center}\caption[]{
Dependence of $\Peff(\beta)$ on the value of $g$ (in units of $k_1\cs^2$).
Note that the depth of the minimum decreases with increasing gravity.
}\label{qp_dpii_grav_x}\end{figure}

We mention in passing the effect of changing gravity.
It is clear that increasing gravity enhanced the anisotropy
of the turbulence, which seems to have a reducing effect on
the negative effective magnetic pressure; see \Fig{qp_dpii_grav_x}.
The reason for this is at the moment not well understood.
We emphasize that this effect is connected with $\qp$ and not with
$\qg$ that was introduced in \Eq{qg}.

\subsection{MFS}
\label{MFS}

\begin{figure}[t!]\begin{center}
\includegraphics[width=\columnwidth]{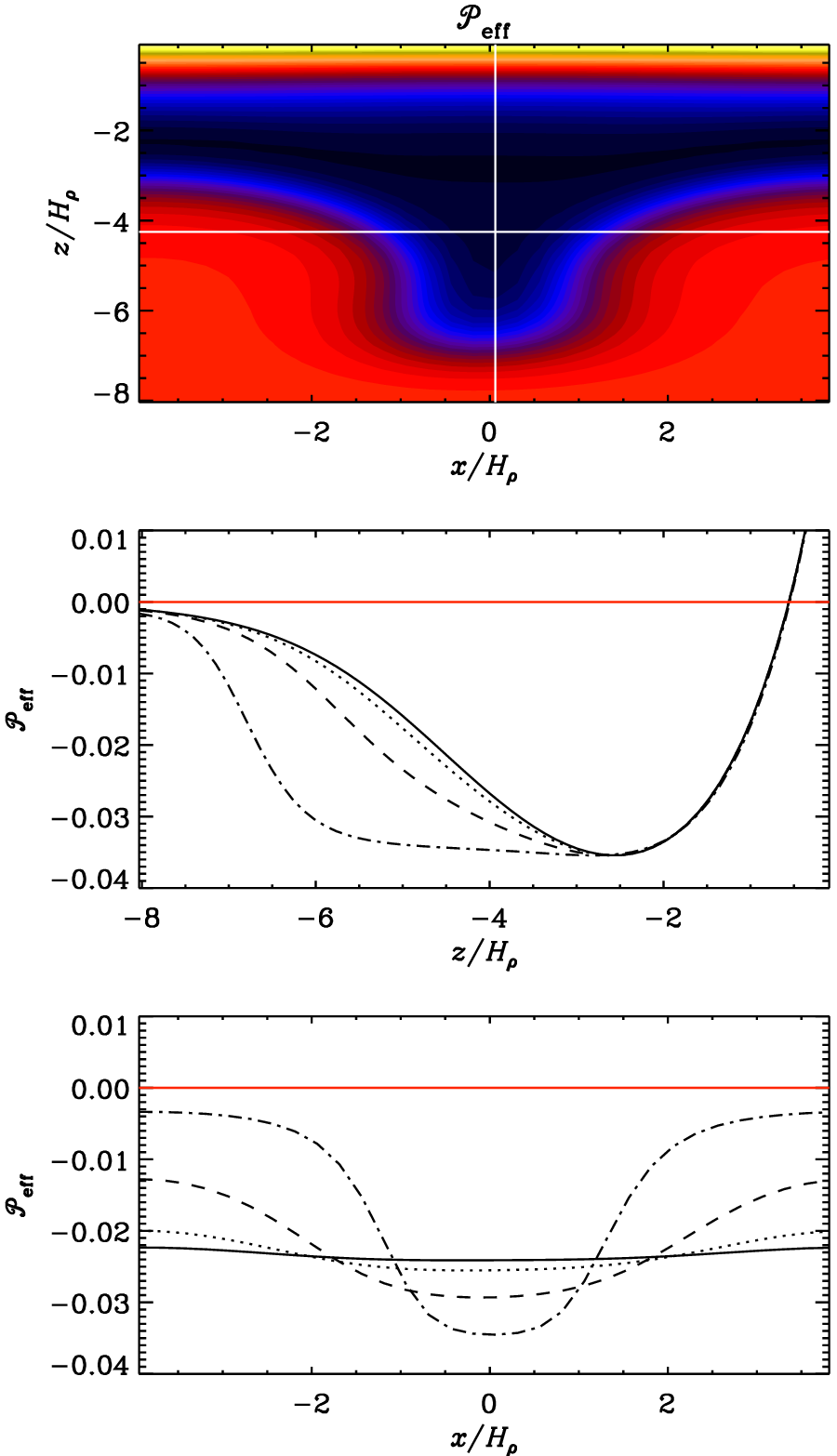}
\end{center}\caption[]{
Representation of $\Peff$ in the $xz$ plane using a MFS with
$\etatz/\urms H_\rho=10^{-2}$, corresponding to $\kf H_\rho\approx30$,
$B_0/\Beqz=0.4$, $\betastar=0.32$ and $\betap=0.05$.

}\label{MF_Pb}\end{figure}

Next, let us investigate the spatial variations of $\Peff$ in a
corresponding MFS.
We use the parameters $\betastar$ and $\betap$ that are appropriate
in the regime investigated in the DNS above, namely
$\etatz/\urms H_\rho=10^{-2}$, corresponding to $\kf H_\rho\approx30$,
$B_0/\Beqz=0.4$, $\betastar=0.32$ and $\betap=0.05$.
The result is shown in \Fig{MF_Pb}, where we plot in the upper panel
the $xz$ dependence of $\Peff$.
Since the domain is periodic in the $x$ direction, we were able to shift
the position of the minimum such that it lies approximately at $x=0$.
The white vertical line near $x=0$ and the horizontal white line near
$z/H_\rho=-4.3$ indicate positions along which we plot in the next two
panels $\Peff$ at three different times.

Initially, the minimum of $\Peff(\beta)$ occurs at the height
$z/H_\rho\approx-2.5$, but at later times the minimum broadens
and we have $\Peff\approx-0.035$ in the range $-5.5<z/H_\rho<2.5$.
In the last panel of \Fig{MF_Pb} we show that the horizontal
extent of the structure becomes narrower and more concentrated
as time goes on.

\section{Conclusions}

The present results have shown that NEMPI tends to develop sharp
structures in the course of its nonlinear evolution.
This becomes particularly clear from the MFS presented in \Sec{MFS}.
The results of \Sec{DNS} suggest that this might have consequences
of an intensification of NEMPI with increasing $|\meanJJ|$, as was
demonstrated using DNS.
At present it is not clear what is the appropriate parameterization
of this effect.
One possibility is that the $\meanJJ$ dependence enters in the same
way as the $\meanBB$ dependence, i.e.,
$\overline\Pi_{xx}=-\half(1+\meanJ^2/k_J^2\Beq^2)\qp(\beta)\,\meanB^2$,
where we treat $k_J$ as a free parameter, although this might be a naive
expectation given the small number of data points and experiments performed.

The present results are just a first attempt in going beyond the
simple representation of the turbulent stress in terms if the
mean field along.
Other important terms include combinations with gravity as well as
anisotropies of the form $\meanJ_i\meanJ_j$.
Furthermore, if there is helicity, one could construct contributions to
the stress tensor using products of the pseudo-tensors $\meanJ_i\meanB_j$
and $\meanJ_j\meanB_i$ with the kinetic or magnetic helicity.
Such a construction obeys the fact that the Reynolds and Maxwell tensors
are proper tensors.
Such pseudo-tensors might play a role in the solar dynamo where the
$\alpha$ effect is believed to play an important role.
However, nothing is known about the importance or the sign of such effects.
It would thus be desirable to have an accurate method that allows one to
determine the relevant turbulent transport coefficients.

\section*{Acknowledgements}

We thank K.-H.\ R\"{a}dler who suggested to take into account the
effect of the current density to NEMPI.
We are grateful for the allocation of computing resources provided by the
Swedish National Allocations Committee at the Center for
Parallel Computers at the Royal Institute of Technology in Stockholm.
This work was supported in part by
the European Research Council under the AstroDyn Research Project
227952, the Swedish Research Council grant 621-2011-5076 (AB),
by COST Action MP0806,
by the European Research Council under the Atmospheric Research Project No.\ 227915
and by a grant from the Government of the Russian Federation under
contract No.\ 11.G34.31.0048 (NK,IR).

\bibliographystyle{aa}

\begin{thebibliography}{100}

\bibitem[Brandenburg(2005)]{B05}
Brandenburg A\yapj{2005}{625}{539}

\bibitem[Brandenburg and Subramanian(2005)]{BS05}
Brandenburg A and Subramanian K\yjour{2005}{Phys.\ Rep.}{417}{1}

\bibitem[Brandenburg \etal(2010)]{BKR10}
Brandenburg A, Kleeorin N and Rogachevskii I\yan{2010}{331}{5}

\bibitem[Brandenburg \etal(2011)]{BKKMR}
Brandenburg A, Kemel K, Kleeorin N, Mitra D and Rogachevskii I\yapj{2011}{740}{L50}

\bibitem[Brandenburg \etal(2012)]{BKKR12}
Brandenburg A, Kemel K, Kleeorin N and Rogachevskii I\yapj{2012}{749}{179}

\bibitem[Cally \etal(2003)]{Cally}
Cally P S, Dikpati M and Gilman P A\yapj{2003}{582}{1190}

\bibitem[Gilman and Dikpati(2000)]{GD00}
Gilman P A and Dikpati M\yapj{2000}{528}{552}

\bibitem[K\"apyl\"a \etal(2012a)]{KMB12}
K\"apyl\"a P J, Mantere M J and Brandenburg A\papjl{2012a}{1205.4719}

\bibitem[K\"apyl\"a \etal(2012b)]{KBKMR12}
K\"apyl\"a P J, Brandenburg A, Kleeorin N, Mantere M J and Rogachevskii I\ymn{2012b}{422}{2465}

\bibitem[Kemel \etal(2012a)]{KBKR12}
Kemel, K., Brandenburg, A., Kleeorin, N. and Rogachevskii, I.\yan{2012}{333}{95}

\bibitem[Kemel \etal(2012b)]{KBKMR12a}
Kemel K, Brandenburg A, Kleeorin N, Mitra D and Rogachevskii I\psph{2012b}
{DOI:10.1007/s11207-012-0031-8}{arXiv:1203.1232}

\bibitem[Kemel \etal(2012c)]{KBKMR12b}
Kemel K, Brandenburg A, Kleeorin N, Mitra D and Rogachevskii I\psph{2012c}
{DOI:10.1007/s11207-012-9949-0}{arXiv:1112.0279}

\bibitem[Kitchatinov \& Mazur(2000)]{KM00}
Kitchatinov LL, Mazur MV\ysph{2000}{191}{325}

\bibitem[Kleeorin \etal(1989)]{KRR89}
Kleeorin N, Rogachevskii I and Ruzmaikin A\ysovl{1989}{15}{274}

\bibitem[Kleeorin \etal(1990)]{KRR90}
Kleeorin N, Rogachevskii I and Ruzmaikin A\yjetp{1990}{70}{878}

\bibitem[Kleeorin \etal(1993)]{KMR93}
Kleeorin N, Mond M and Rogachevskii I\ypfb{1993}{5}{4128}

\bibitem[Kleeorin \etal(1996)]{KMR96}
Kleeorin N, Mond M and Rogachevskii I\yana{1996}{307}{293}

\bibitem[Kleeorin and Rogachevskii(1994)]{KR94}
Kleeorin N and Rogachevskii I\ypre{1994}{50}{2716}

\bibitem[Krause and R\"adler(1980)]{KR80}
Krause F and R\"adler K-H\ybook{1980}
{Mean-field magneto\-hydro\-dy\-na\-mics and dynamo theory}
{Pergamon Press, Oxford}

\bibitem[Losada(2012)]{Los12}
Losada I R, Brandenburg A, Kleeorin N, Mitra D and Rogachevskii I\sana{2012},
arXiv:1207.5392

\bibitem[Moffatt(1978)]{Mof78}
Moffatt H K\ybook{1978}
{Magnetic field generation in electrically conducting fluids}
{Cambridge University Press, Cambridge}

\bibitem[Parker(1979)]{Par79}
Parker E N\ybook{1979}{Cosmical magnetic fields}{Oxford University Press, New York}

\bibitem[Parfrey and Menou(2007)]{PM07}
Parfrey K P and Menou K\yapj{2007}{667}{L207}

\bibitem[Rogachevskii \& Kleeorin(2007)]{RK07}
Rogachevskii I and Kleeorin N\ypre{2007}{76}{056307}

\bibitem[Stein and Nordlund(2012)]{SN12}
Stein R F and Nordlund \AA\yapj{2012}{753}{L13}

\bibitem[Zeldovich \etal(1983)]{ZRS83}
Zeldovich Ya B, Ruzmaikin A A and Sokoloff D D\ybook{1983}
{Magnetic fields in astrophysics}{Gordon \& Breach, New York}

\end{thebibliography}

\end{document}